\documentclass{optica-article}

\usepackage{graphicx}
\usepackage{subcaption}
\usepackage{multirow}
\usepackage{xcolor}
\usepackage[export]{adjustbox}

\journal{opticajournal} 

\articletype{Research Article}

\usepackage{lineno}

\begin{document}

\title{\textcolor{black}{Weak Kerr Nonlinearity Boosts the Performance of Frequency-Multiplexed Photonic Extreme Learning Machines: A Multifaceted Approach}}

\author{Marina Zajnulina,\authormark{1} Alessandro Lupo,\authormark{2,*} and Serge Massar\authormark{2}}

\address{
\authormark{1} Multitel Innovation Centre, Rue Pierre et Marie Curie 2, 7000 Mons, Belgium

\authormark{2}Laboratoire d'Information Quantique CP224, Université libre de Bruxelles (ULB), Av. F. D. Roosevelt 50, 1050 Bruxelles, Belgium

}


\email{\authormark{*}alessandro.lupo@ulb.be} 

\begin{abstract*}
\textcolor{black}{We provide a theoretical, numerical, and experimental investigation of the Kerr nonlinearity impact on the performance of a frequency-multiplexed Extreme Learning Machine (ELM). In such ELM, the neuron signals are encoded in the lines of a frequency comb. The Kerr nonlinearity facilitates the randomized neuron connections allowing for efficient information mixing. A programmable spectral filter applies the output weights. The system operates in a continuous-wave regime. Even at low input peak powers, the resulting weak Kerr nonlinearity is sufficient to significantly boost the performance on several tasks. This boost already arises when one uses only the very small Kerr nonlinearity present in a 20-meter long erbium-doped fiber amplifier. In contrast, a subsequent propagation in 540 meters of a single-mode fiber improves the performance only slightly, whereas additional information mixing with a phase modulator does not result in a further improvement at all. We introduce a model to show that, in frequency-multiplexed ELMs, the Kerr nonlinearity mixes information via four-wave mixing, rather than via self- or cross-phase modulation. At low powers, this effect is quartic in the comb-line amplitudes. Numerical simulations validate our experimental results and interpretation.}
\end{abstract*}

\section{Introduction}

Artificial neural networks (ANNs) are brain-inspired algorithms that process information by propagating it \textcolor{black}{through a series} of weighted connections and nonlinear activation functions. ANNs are usually implemented in electronic digital computers, mainly through matrix multiplications with a subsequent application of nonlinear functions. Due to the separation between computing and memory units, the classical von Neumann computer architecture suffers a severe performance bottleneck when executing algorithms \textcolor{black}{that} massively \textcolor{black}{depend} on memory access, \textcolor{black}{such as ANNs} \cite{Beyond2020}. \textcolor{black}{For this reason, ANN tasks are usually executed on auxiliary
hardware (GPUs or AI accelerators) } \cite{Reuther2019}. \textcolor{black}{In these devices,} the information \textcolor{black}{flows through an} artificial structure \textcolor{black}{of hard-wired networks and/or logic gates}. \textcolor{black}{In opposition}, the biological brain exploits the \textcolor{black}{complex} dynamics of its organic \textcolor{black}{material}. This inspires alternative approaches to computing that \textcolor{black}{harvest} the dynamics of \textcolor{black}{the} physical substrate \textcolor{black}{they are built on,} \textcolor{black}{leading to the implementation of} analog ``physical computers'' \cite{Markovi2020, Nakajima2020}.

Extreme Learning Machines (ELMs) are feed-forward ANNs with fixed, i.e.\ not trained, connections between the input and hidden layers, while only the \textcolor{black}{output-layer} connections undergo a training procedure \textcolor{black}{to find optimal weights}\cite{huang2006extreme, huang2015extreme}. The \textcolor{black}{output-layer} training can be done using non-iterative algorithms such as linear regression, which is by far less energy- and time-consuming than the backpropagation algorithm usually employed for training \textcolor{black}{of} fully connected networks. It has been proven that training only the output connections is enough to guarantee the universal approximation property of ELMs \cite{Huang2006}\textcolor{black}{, although this is only a necessary (but not sufficient) condition for good performance.}
Since there is no requirement for \textcolor{black}{internal-connections} tuning, ELMs are particularly suitable for implementation on physical substrates. In such physical implementations, the substrate nonlinearly mixes the input signals and therefore plays the role of connectivity between the input and hidden layers. \textcolor{black}{Thus, Ref.~\cite{Marcucci2020} reports 
a theoretical study of the computing capability of nonlinear waves such as rogue waves, dispersive shocks, and solitons.}  

Photonic systems constitute an interesting substrate for analog computation. They \textcolor{black}{are not only able to provide different types of nonlinearity for information processing, but also exhibit} multiple degrees of freedom \textcolor{black}{for information encoding, as well as} the potential for low-loss, \textcolor{black}{ultra-fast, and high-bandwidth data processing and throughput}\cite{Shastri2021}. For example, photonic ELMs have been implemented based on propagation through scattering media \cite{saade2016, Pierangeli21} \textcolor{black}{and} \textcolor{black}{multimode fibers \cite{Tein2021}}, \textcolor{black}{as well as} in delay-based schemes \cite{ortin2015unified}, \textcolor{black}{and, in general, in any system whose dynamics is described by 
 the Nonlinear Schr\"odinger equation (NLS) \cite{Marcucci2020}}.

Using the frequency of light as a degree of freedom is a natural solution for processing large numbers of signals in parallel. Frequency-multiplexed modes can be generated and manipulated by employing reliable and efficient optical components, such as phase and intensity modulators, programmable spectral filters, and other devices for wavelength-division multiplexing (WDM). All these components can be integrated on a photonic chip and are already massively employed for telecommunication. Recently, various frequency-multiplexing-based schemes have been used to implement physical ANNs \cite{xu2020, xu2021,feldmann2021, Lupo2021, Lupo2022, yildirim2022, Butschek2022,Zhou2022, Sorokina2020, Fischer2023}, where, depending on the experiment, the frequency-multiplexed information is manipulated by wavelength-dependent couplers, \textcolor{black}{or exploiting the dispersion or optical nonlinearities of the material}.

In \textcolor{black}{our} previous work \cite{Lupo2021, Lupo2022}\textcolor{black}{,} we introduced a photonic ELM that uses frequency multiplexing to encode the input information in the lines of an optical frequency comb. The information mixing in the input-to-hidden layer connections was realized via frequency interference induced by periodical phase modulation, giving rise to a transformation linear in the electrical field amplitudes, i.e. \textcolor{black}{we had} a linear mixing of input signals. \textcolor{black}{The required nonlinearity was quadratic. It was provided by a photodiode that measured the intensities of the comb lines.}

Here, we \textcolor{black}{exploit the Kerr nonlinearity as} a fundamentally different information processing \textcolor{black}{mechanism. It arises in} single-mode optical fibers (erbium-doped and standard telecom fibers) \textcolor{black}{deployed in the experiment}. \textcolor{black}{Thus, we} use a low-power continuous wave (CW) operation in an approximately $20$-meter-long erbium-doped fiber optionally followed by a $540$-meter-long standard telecom fiber.
 
\textcolor{black}{
We demonstrate our system on four benchmark tasks, consisting of classification and regression problems. We also compare our system to software-based ELMs with the same number of nodes, obtaining similar prediction accuracy. Finally, we present an analytical model that shows that the information mixing in our system occurs due to four-wave mixing (FWM); dispersion effects as well as self- and cross-phase modulation play no role. In the low-power regime, the Kerr effect introduces a quartic nonlinearity (as opposed to quadratic nonlinearity from our previous work).}

We find that using the Kerr nonlinearity \textcolor{black}{generated in a piece of only 20-meter-long erbium-doped fiber amplifier (EDFA) is sufficient to significantly boost the performance of our ELM on several tasks. Subsequent propagation in 540 meters of a single-mode fiber improves the performance only slightly. Linear information mixing with a following phase modulator does not result in a further improvement at all. As the measured EDFA output is only in the rage of 0.5W, we deal with a weak overall Kerr nonlinearity in the system. This means that a weak Kerr nonlinearity is sufficient for effective and efficient information processing.}

\textcolor{black}{
The sufficiency of a weak Kerr nonlinearity is a counterintuitive result opposing the general idea that a high nonlinearity is required for photonic neuromorphic computing (cf. \cite{Marcucci2020}). In fact, Refs.~\cite{Fischer2023} and~\cite{Zhou2022} report using highly nonlinear fibres (HNLF) and femto-second (pulsed) lasers as sources to facilitate  information processing. 
Highly nonlinear regimes require high peak power pulsed lasers and/or special fibers, effectively reducing the applicability of the approach. Using a weak Kerr nonlinearity
makes the system easier to implement.
We note that a successful usage of a weak Kerr nonlinearity was also studied in Ref.~\cite{Pauwels2019} in the context of Reservoir Computing and in Refs.~\cite{Kryzhanovsky2003, kryzhanovsky2001parametric} in the context of Hopfield networks, showing the generalization potential of this concept. In addition, fiber-based schemes that deploy weak Kerr nonlinearities would be easier to transfer on semiconductor chips than the schemes that operate in highly nonlinear regimes, as they require less optical power and yield less thermal dissipation.}

\textcolor{black}{
The paper is structured as follows. Sec.~\ref{sec:principles} describes the principles of a generic ELM algorithm, the impact of the Kerr nonlinearity and frequency-domain interference; Sec.~\ref{sec:methods} describes the experimental setup, as well as its operating procedures; Sec.~\ref{sec:results} contains the experimental results and their discussion; Sec.~\ref{sec:conclusion} contains the conclusions. App.~\ref{appendix:A} depicts the low-power analytical model for information mixing driven by optical Kerr effect. App.~\ref{appendix:B} reports a summary of the benchmarking results in every tested configuration.}


\section{Principles}
\label{sec:principles}
\subsection{Extreme Learning Machine Algorithm}
An Extreme Learning Machine (ELM) is a feed-forward neural network where only the output weights are trained, while the other connections are fixed and \textcolor{black}{remain untrained} \cite{huang2006extreme, huang2015extreme}. We define $N^\mathit{in},$ $N^\mathit{hidden},$ and $N^\mathit{out}$ \textcolor{black}{as} the numbers of the input, hidden, and output \textcolor{black}{neurons,} respectively. $\mathbf{u}$, $\mathbf{h}$, and $\mathbf{y}$ \textcolor{black}{are} the column vectors containing the \textcolor{black}{corresponding neuron values.} \textcolor{black}{In general,} a single-hidden--layer ELM is described by the following two equations:
\begin{align}
    \mathbf{h} &= f_\mathit{hidden}\left(\mathbf{W}^\mathit{in}\cdot f_\mathit{in}\left(\mathbf{u}\right)\right),\label{eq:hidden_layer}\\
    \mathbf{y} &= \mathbf{W}^\mathit{out}\cdot\mathbf{h}, \label{eq:output_layer}
\end{align}
where $f_\mathit{in}$ and $f_\mathit{hidden}$ are the nonlinear activation functions of the input and hidden layers \textcolor{black}{(they act elementwise),} $\mathbf{W}^\mathit{in}$ is a $N_\mathit{hidden}\times N_\mathit{in}$ matrix of the input-to-hidden layer weights, and $\mathbf{W}^\mathit{out}$ is a $N^\mathit{output}\times N^\mathit{hidden}$ matrix of the hidden-to-output layer weights. The matrix $\mathbf{W}^\mathit{in}$ is not \textcolor{black}{trained,} but rather selected at random.  $\mathbf{W}^\mathit{out}$ is trained employing ridge regression to achieve an output $\mathbf{y}$ \textcolor{black}{that approximates the true value.}

In physical implementations of ELMs, the transformation  $\mathbf{h}( \mathbf{u})$ between the input and the hidden layer is fixed by the physical system and may therefore take a form different from Eq.\ \eqref{eq:hidden_layer}. However it \textcolor{black}{is important} for good operation that
 $\mathbf{h}( \mathbf{u})$ is a nonlinear mapping from the $N^\mathit{in}$ input nodes to a larger number of hidden nodes $N^\mathit{hidden} \gg N^\mathit{in}$. \textcolor{black}{In the context of ANNs, this mapping corresponds to the mapping of the input to a higher-dimensional space which generally allows for a better data separability.}

\subsection{Extreme Learning Machine based on Frequency Multiplexing}
\label{sec:ELM_FM}

\paragraph{General Framework.}

\textcolor{black}{As in our previous publications  (cf.~\cite{Lupo2021,Lupo2022})}, we use an optical frequency comb \textcolor{black}{with central frequency $\omega_0$ and line spacing $\Omega$} to encode and process the input data. \textcolor{black}{The} k-th line in the comb reads \textcolor{black}{accordingly} as $\omega_{k} = (\omega_0 + k\Omega)$, where $k$ \textcolor{black}{is a positive or negative integer} representing frequencies higher \textcolor{black}{or} lower than $\omega_{0}$. 

\textcolor{black}{We encode the input data $\mathbf{u}$ (cf. Eq.~\ref{eq:hidden_layer}) in the comb-line amplitudes $x^\textit{in}_{k}$ by applying attenuations or phase delays with corresponding values. More details on data encoding can be found in Sec.~\ref{sec:procedure}. With encoded input data, the time-domain representation of the comb reads as:}
\begin{equation}
      E^\textit{in}(t)=\sum_{k} x^\textit{in}_{k}  (\mathbf{u} ) e^{-i\left(\omega_0+k\Omega\right)t}\ ,
\label{eq:opt_sig}
\end{equation}
where $t$ represents the time, $k$ is the comb line index with $k=0$ corresponding to the central line, \textcolor{black}{and} $x^\textit{in}_{k}$ is the k-th line complex amplitude. Since the frequency comb has a limited spectral \textcolor{black}{range}, the elements of the sum in Eq.\ \eqref{eq:opt_sig} vanish for \textcolor{black}{a} sufficiently large $\vert k \vert$.

After propagation \textcolor{black}{through the system,} the electric field is transformed into the output field:
\begin{equation}
      E^\textit{out}(t)=\sum_{k} x^\textit{out}_{k}  (\mathbf{u} ) e^{-i\left(\omega_0+k\Omega\right)t}\ ,
\label{eq:opt_sig_out}
\end{equation}
where the output amplitudes are functions of the input amplitudes:
\begin{equation}
{x}^\textit{out}_k  =f _k(\mathbf{x}^\textit{in})\ .
\label{eq:Xout_generic}
\end{equation}
In our previous \textcolor{black}{work,} the transformation Eq.~\eqref{eq:Xout_generic} was linear and realized by a phase modulator \cite{Lupo2021}. \textcolor{black}{Here, the transformation is facilitated by Kerr nonlinearity as it will be shown in the further course of our study.}

\textcolor{black}{We obtain the ELM output} by taking a linear combination of the comb line intensities:
\begin{equation}
y_\textit{j} = \sum_k W^\textit{out}_{jk} \vert x^\textit{out}_{k} \vert^2\ .
\label{eq:Xout}
\end{equation}
where $\mathbf{W}^\textit{out} $ is trained by ridge regression. Note that the operation Eq.\ \eqref{eq:Xout} can be realised optically, as reported in \cite{Lupo2021} and described below. In summary, the vector of output comb intensities \textcolor{black}{has the meaning of a} hidden layer
$\mathbf{h} =\vert \mathbf{ x^\textit{out}} \vert^2$, with Eqs. (\ref{eq:opt_sig}, \ref{eq:opt_sig_out}, \ref{eq:Xout_generic}, \ref{eq:Xout}) \textcolor{black}{corresponding to} Eqs. (\ref{eq:hidden_layer}, \ref{eq:output_layer}).

\paragraph{\textcolor{black}{Fiber Kerr Nonlinearity for Information Mixing.}}
\begin{figure}
\centering
\begin{subfigure}[b]{0.8\textwidth}
    \includegraphics[width=\linewidth]{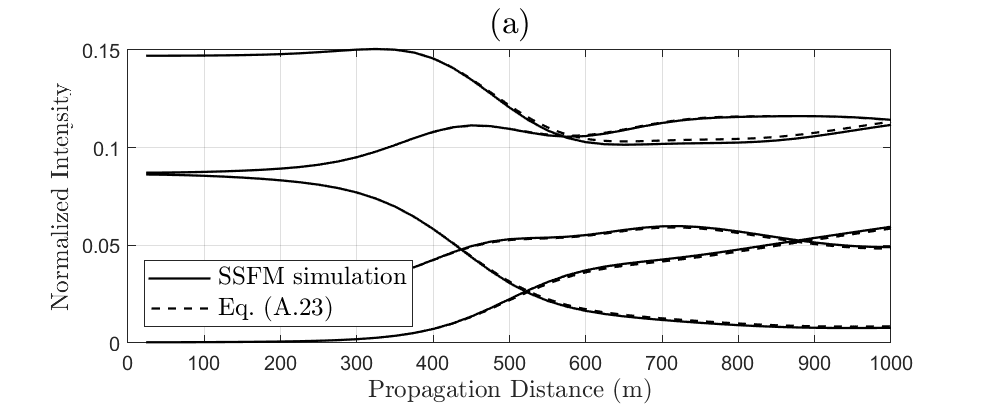}    
\end{subfigure}

\vspace{15 px}

\begin{subfigure}[b]{0.8\textwidth}
    \includegraphics[width=\linewidth]{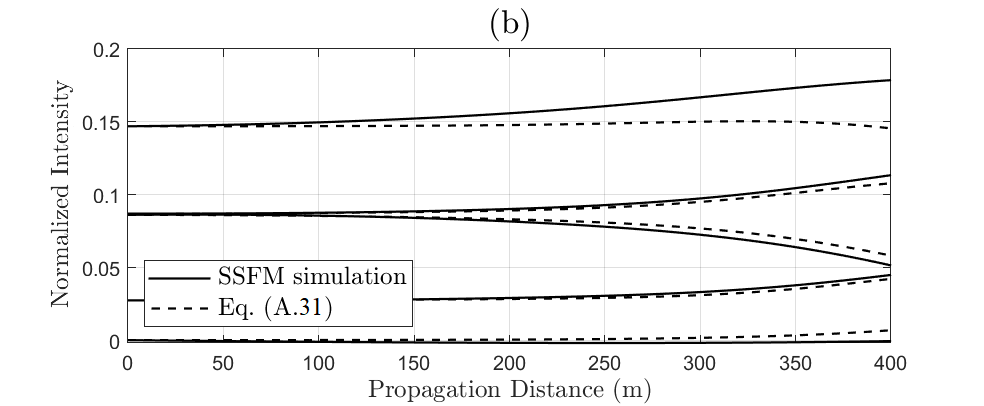}  
\end{subfigure}

\vspace{15 px}

\begin{subfigure}[b]{0.8\textwidth}
    \includegraphics[width=\linewidth]{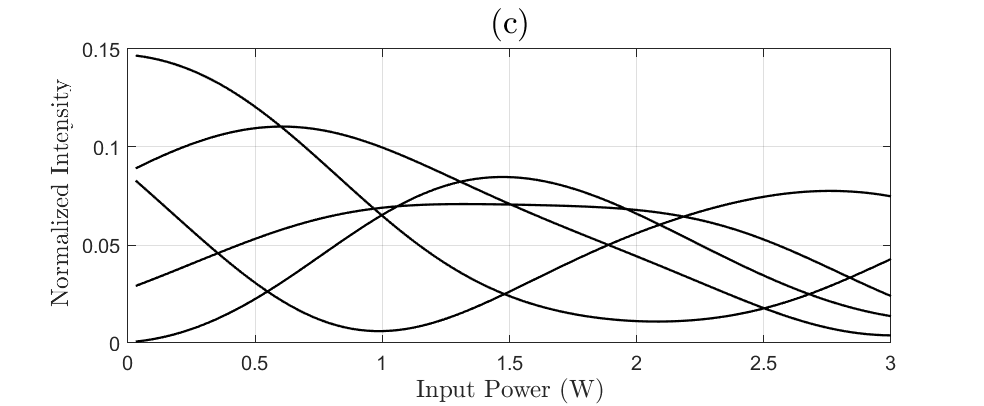}  
\end{subfigure}

  \caption{
  Evolution of \textcolor{black}{five most powerful comb-line intensities normalized by comb total power with fiber parameters taken from} Sec.\ \ref{sec:setup}.\\\hspace{\textwidth} 
  (a) \textcolor{black}{Normalized comb-line intensities as a function of fiber propagation length for a fixed input power of $0.5$ W. Solid lines are obtained by integration of full Nonlinear Schr\"odinger equation via split-step Fourier method (SSFM). Dashed lines show intensities obtained via integration of comb-amplitude equation derived explicitly for our Extreme Learning Machine (Eq.\ \eqref{Eq:NLS2} in App. \ref{appendix:A}). The two methods should be equivalent and the slight diffference is probably due to numerical precision.} \\\hspace{\textwidth}
  (b) Same as (a), but dashed lines are produced \textcolor{black}{by integration of Eq.~\ \eqref{Eq:DeltaIk} obtained via perturbing the Nonlinear Schr\"odinger equation and taking the terms in the order of $\gamma$ (App.~\ref{appendix:A}). 
The figure shows that the linearized approximation of Eq. \eqref{Eq:DeltaIk} is valid for approximately $100$ meters of propagation.  
} \\\hspace{\textwidth}
  (c) \textcolor{black}{Normalized comb-line intensities as a function of a function of the input power for a fixed fiber propagation length of $L=540$ m obtained via integration of the Nonlinear Schr\"odinger equation.}}
  \label{fig:Simul}
\end{figure}

\textcolor{black}{The input transforms to the output as the comb propagates} through an optical fiber of length $L$ (cf. Eq.~\eqref{eq:Xout_generic}). During this propagation, the \textcolor{black}{slowly varying electric-field amplitude $E(z, t)$ formulated in the co-moving frame} undergoes linear (dispersive) and nonlinear (Kerr-effect-based) transformations described \textcolor{black}{by NLS} \cite{Marcucci2020, Agrawal}:
\begin{equation}
    \frac{\partial E}{\partial z} = -i\frac{\beta_{2}}{2}\frac{\partial^{2}E}{\partial t^{2}} + i\gamma|E|^{2}E - \frac{\alpha}{2}{E},
    \label{eq:GLNS}
\end{equation}
where $E(z,t)$ is the \textcolor{black}{electric-field} amplitude at position $z$ \textcolor{black}{and} time $t$, $\beta_{2}$ \textcolor{black}{is} the group-velocity dispersion, $\gamma$ is the nonlinear coefficient that describes the intensity-dependent Kerr effect, and $\alpha$ \textcolor{black}{are the optical losses (with $\alpha<0$ corresponding to a gain, as in the case of an EDFA).} 

The electric field at the output of the optical fiber reads as:
\begin{equation}
E(L,t)=    \sum_{k} x^\textit{Kerr}_\textit{k}e^{-i(\omega_0 +k\Omega) t}.
\label{eq:opt_sig_Kerr}
\end{equation}
where $x^\textit{Kerr}_{k}$ are \textcolor{black}{comb-line amplitudes transformed by Kerr nonlinearity. The corresponding $j$-th output of the ELM  takes the form:}
\begin{equation}
y_\textit{j}=    \sum_\textit{k}  W^{out}_\textit{jk} \vert  x^\textit{Kerr}_\textit{k}\vert^2\ .
\label{eq:out_Kerr}
\end{equation}

\textcolor{black}{
Eq.~\ref{eq:GLNS} is not integrable. Therefore, we cannot provide an analytical expression for the dependence between the input $(x^\textit{in}_{k})$ and output $(x^\textit{Kerr}_\textit{k})$ comb lines. However, we derived a model for the Kerr-based comb-line transformation assuming a weak Kerr nonlinearity, i.e. a nonlinearity of the first order in $\gamma$. More details on this model can be found in App.~\ref{appendix:A}.}

Our model shows that 
\begin{itemize}
    \item Kerr-nonlinearity-driven FWM is the only process that is responsible for information mixing in our ELM. It affects the intensities of the comb lines, which is detected by ELM's readout layer. In contrast, self- and cross-phase modulation do not contribute to information mixing as these effects act on the phases of $x^\textit{Kerr}_\textit{k}$ and the phase information is lost when the comb line intensities are read out.
    \item 
The change in the comb line intensities, i.e. the information mixing, depends quartically on comb line amplitudes If we further suppose that the porpagation distance $L$ is very small, so that dispersion and loss can be neglected, then the output intensities take the form
\begin{equation}
\vert  x^\textit{Kerr}_{k}\vert^2 
\approx
\vert  x^\textit{in}_{k}\vert^2 
-  \gamma  L \operatorname{Im} 
\left(
\sum_{\substack{l-m+n=k \\ n\neq k ; l\neq k}}
 x_l^\textit{in} \overline{ x_m^\textit{in}} x_{n}^\textit{in} \overline {x_k^\textit{in}}  
 \right)\ .
 \label{eq:KerrEffect}
 \end{equation}
\end{itemize}

The average power in our system is $P_0\approx0.5$ W \textcolor{black}{which leads to invoking only a weak Kerr nonlinearity.} A numerical simulation of the comb evolution as a function of input power and propagation length is reported in Fig.\ \ref{fig:Simul}.
The figure shows that the comb line intensities are not linearly varying over the relevant power range, therefore showing that the first order expansion in $\gamma$ used in Eq.~\ref{eq:KerrEffect} does not describe adequately the experimental situation, although it does describe adequately the initial propagation. Moreover, the neglect of the dispersion  is not justified experimentally. Nevertheless, Eq.~\ref{eq:KerrEffect}, and its generalizations given in App. \ref{appendix:A}  provide a rough idea of how the Kerr effect mixes information in the fiber.

\paragraph{\textcolor{black}{Phase Modulation for Information Mixing.}}
\textcolor{black}{To build a bridge between our current and previous work, we tested whether a phase modulation applied to the output of a fiber can further increase the performance of our ELM. As described in Refs.~\cite{Lupo2021,Lupo2022}, the phase modulation is performed at comb-line spacing frequency $\Omega.$ It induces frequency-domain interference between the lines and, thus, constitutes a possible mechanism for effective information mixing. Contrary to Kerr-nonlinearity mixing, the mixing process by phase modulation is a linear process.} 

We define $x^\textit{PM}_k$ \textcolor{black}{as the comb-line amplitudes at the output of a phase modulator} \cite{Lupo2021}:
\begin{equation}
  \sum_kx^\textit{PM}_k e^{-ik(\omega_0 +k\Omega)t} = \sum_{k} x^\textit{Kerr}_{k}e^{-i(\omega_0 +k\Omega) t}e^{-im\cos{(\Omega t)}}\ .
  \label{Eq:PM-A}
\end{equation}
where the time-dependent phase $m\cos{(\Omega t)}$ is the result of the phase modulation, with $m$ being \textcolor{black}{the} coefficient defining the intensity of the phase modulation. Using the Jacobi-Anger expansion $e^{-im\cos{(\Omega t)}} = \sum_l  i^l J_l(m) e^{-i(k+l)\Omega t},$ \textcolor{black}{we can rewrite modulated comb amplitudes as Bessel functions:}
\begin{equation}
    \label{eq:pm}
    x^\textit{PM}_k = \sum_l x^\textit{Kerr}_li^{k-l}J_{k-l}(m)
\end{equation}
with $J_l$ \textcolor{black}{being} $l$-th order Bessel function of the first kind. \textcolor{black}{In the current setup, we use} $m\approx2$. 

\section{Methods}\label{sec:methods}

\subsection{Experimental Setup}
\label{sec:setup}

\begin{figure}
\centering
\includegraphics[width=0.98\textwidth, frame]{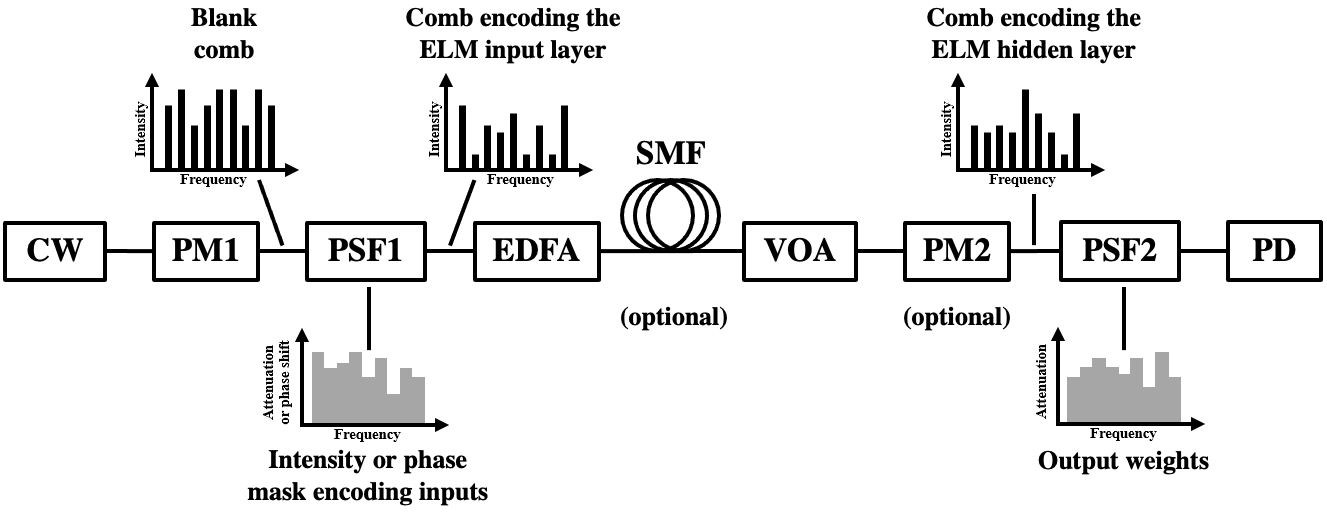}
\caption{Experimental \textcolor{black}{setup} of a fiber-based Extreme Learning Machine. CW: \textcolor{black}{continuous-wave} laser; PM: phase modulator; EDFA: erbium-doped fiber amplifier; PSF: programmable spectral filter SMF: single-mode fiber; VOA: variable optical attenuator; PD: photodiode. The source of the radio frequency (RF) signal driving PM1 and (optionally) PM2 is not represented.}\label{fig:ExpSetup}
\end{figure}

Fig.~\ref{fig:ExpSetup} shows the experimental setup. The input frequency comb is generated by phase modulating \textcolor{black}{the output of a continuous-wave laser (CW) with a lithium-niobate} phase modulator (PM1). The CW wavelength of $\lambda=1554.6\;\mathrm{nm}$ \textcolor{black}{corresponds to comb's central wavelength.}
PM1 is driven by a radio-frequency (RF) signal at frequency $\Omega \approx 17\;\mathrm{GHz}$ and power $P1\approx30$ dBm. $\Omega$ determines the comb line spacing, while $P1$ determines the number of comb lines. The current setup generates approximately 25 comb lines spanning a spectrum of $\sim3$ nm. Note that at the high RF power used to drive PM1, the transformations Eqs.\ (\ref{Eq:PM-A}, \ref{eq:pm}) do not describe exactly the effect of PM1, probably because of distortions in the sinusoidal RF signal (second-order corrections could be applied to account for this effect, see \cite{Lupo2021} for details).

\textcolor{black}{We encode the input data using} a programmable spectral filter (PSF1, \textit{Coherent II-VI Waveshaper}). \textcolor{black}{A PSF wavelength-multiplexes the comb lines, applies variable attenuation and/or phase shifts to them, and recombines them to a single output comb. Using PSF1, we encode the input data by applying either attenuation or a phase delay to each comb line (Sec.\ \ref{sec:procedure})}.

A series of two erbium-doped-fiber amplifiers (EDFA, \textit{Keopsys} models \textit{CEFA-C-PB-LP} and \textit{CEFA-C-BO-HP}) increase the input comb power up to $500$ mW (starting from a power of approximately $10$ mW). The characteristics of the erbium-doped fibers are unknown, as the design of the employed devices is not publicly available. \textcolor{black}{Using openly available data sheets of other EDFA manufactures (\textit{Corning ER}) and assuming that the characteristics of the built-in amplifying fibers would be similar, we estimate the following parameters for amplifiers we use: the total amplifying fiber length is $20$ meters, the dispersion parameter $@1550$ nm is $\approx 18$ ps/nm$\cdot$km corresponding to $\beta_2 \approx -23$ ps$^2$/km, whereas the nonlinear coefficient $\gamma$ is $\approx 6.3$ W$^{-1}$  km$^{-1}.$ Accordingly, the} dispersion-induced phase shift between two neighboring comb lines is $\left(\beta_2 \Omega^2 L/2\right)_\textit{EDFA} \approx 0.3$ \textcolor{black}{rad, whereas} the nonlinear phase shift is $\left(\gamma \int_0^L P(z) dz\right)_\textit{EDFA}\approx 0.015$ rad (assuming a constant amplification distributed along the propagation length). Optionally, the comb can also be injected in $L = 540$ m of \textcolor{black}{a} standard telecom single-mode fiber (SMF, \textit{Throlabs SMF-28-J9}). The relevant characteristics of the fiber are: attenuation $\approx 0.18$ dB/km, dispersion parameter $@1550$ nm $\approx 18$ ps/nm$\cdot$km corresponding to $\beta_2 \approx -23$ ps$^2$/km, nonlinear coefficient $\gamma \approx 1.2$ W$^{-1}$  km$^{-1}$. With these parameters, the dispersion-induced phase shift between two neighboring comb lines is $\left(\beta_2 \Omega^2 L/2\right)_\textit{SMF} \approx 9.0$ rad, while the nonlinear phase shift is $\left(\gamma P L\right)_\textit{SMF}\approx0.3$ rad (assuming $0.5$ W of power at the input of the fiber). 

A variable optical attenuator (VOA) placed after the fiber reduces the optical power down to a safe value compatible with the rest of the circuit. Optionally, a second phase modulator (PM2) can be used to introduce an additional transformation of the comb leading to further information mixing. PM2 is driven by the same radio-frequency signal as PM1. However, its output power is $P2\approx20$ dBm due to a different electric amplification. 

The output circuit is composed of a second programmable spectral filter (PSF2, \textit{Coherent II-VI Waveshaper}) and a photodiode (PD). PSF2 can be used either to select individual comb lines, by applying band-pass filters, or to apply an attenuation mask to the comb. In the first case, PD measures the intensity of the selected comb line. In the second case, PD measures the sum of the comb line intensities, each one weighted by the attenuation set on the mask. This operation mode is employed to apply output weights optically and is described in Sec.~\ref{sec:procedure}.

From the ANN perspective, PSF1 encodes the input layer of the ELM. EDFA, SMF (optionally) and PM2 (optionally) provide a mix of input information, thus, generating the hidden layer. PSF2 and PD constitute the output layer. 

\subsection{Working procedure}\label{sec:procedure}
\paragraph{\textcolor{black}{Input Layer.}}
In software implementations of ELMs, the dimension of the input layer $N_\textit{in}$ depends on the task and is, in general, equal to the number of features $N_\textit{feat}$ of the \textcolor{black}{processed} dataset, i.e. $N_\textit{in} = N_\textit{feat}$. In our scheme, \textcolor{black}{the number of input layer neurons} coincides with the number of \textcolor{black}{available} comb lines. \textcolor{black}{This number is governed by the parameters of the experimental setup. However, it} can be bigger than the number of data features, i.e. $N_\textit{in}> N_\textit{feat}.$ \textcolor{black}{We deploy a comb with $N_\textit{in}\approx25$ initial lines, while the number of features depends on the benchmark dataset.} \textcolor{black}{Here, we use datasets with feature numbers of 4 or 15, $N_\textit{feat}\in[4,\,15],$ to test our ELM}. 
\textcolor{black}{
It is therefore necessary to establish a mapping between $N_\textit{feat}$ features to be processed and the $N_\textit{in}$ available input neurons. To do so, we stretch the input-feature vector $\mathbf{x}$ by an integer $d\geq1$ that is optimized for every task (\cite{Lupo2021}). The transformation reads as:}
\begin{equation}
    \mathbf{x}=(x_1,x_2,\ldots,x_{N_{\textit{feat}}-1},x_{N_\textit{feat}})\rightarrow(\underbrace{x_1,x_1,\ldots,x_1}_\text{$d$ times},x_2,\ldots,x_{{N_\textit{feat}}-1},\,\underbrace{x_{N_\textit{feat}},x_{N_\textit{feat}},\ldots,x_{N_\textit{feat}}}_\text{$d$ times}).
\end{equation}
The dimension of the stretched vector \textcolor{black}{can still be} smaller than the number of initially available comb lines. The remaining, i.e. \textcolor{black}{unmodulated}, $N_\textit{in}-N_\textit{feat}\cdot d$ lines are left unmodified and are free to \textcolor{black}{co-propagate through the ELM}. This approach proves beneficial \textcolor{black}{as} compared to full attenuation of the unused comb lines \textcolor{black}{because} it allows for more optical power to propagate in the system, \textcolor{black}{which, in turn,} increases the \textcolor{black}{data-transforming} impact of the Kerr \textcolor{black}{nonlinearity}. Moreover, the initially \textcolor{black}{unmodulated} lines acquire information during the propagation, which contributes to the increase of the dimensionality in the hidden layer. \textcolor{black}{Thus, a trade-off is required: high values of $d$ result in a lower number of unmodulated lines which can contribute to the reduction of the total optical power. On the other hand, low values of $d$ result in a lower number of modulated lines, which can reduce the system's response to the input.} \textcolor{black}{For each dataset used here, we found an optimal value of $d$.}

We tested the encoding of data features in comb line intensities and in phases. When encoding in intensities, the features were converted to attenuations in the range of [-30 dB, 0 dB]. In the case of phase encoding, the features were converted to phase delays in the range of [0 rad, $2\pi$ rad]. Both attenuation and phase delays are set by PSF1.

\paragraph{\textcolor{black}{Output Layer.}}
The hidden layer $\mathbf{h}$ is given by the intensities of the comb lines 
 at the output of VOA (or at the output of PM2 when it is used). The intensity measurements are realized by using PSF2 as a tunable band-pass filter that selects one line at a time \textcolor{black}{to be subsequently} measured by PD. Once $\mathbf{h}$ is recorded, a ridge regression model is employed to retrieve the optimal output weights $\mathbf{W}_\textit{out}$. 

The system allows for two weighting methods \textcolor{black}{that} we call ``digital weighing'' and ``optical weighting''. In the digital weighting mode, the multiplication $\mathbf{W}_\textit{out}\cdot\mathbf{h}$ is performed on a traditional computer. This requires that the full vector of hidden neuron values $\mathbf{h}$ is measured \textcolor{black}{for each inference.} In the optical weighting mode, PSF2 applies attenuation to all comb lines simultaneously, which results in PD measuring the sum of the weighted comb line intensities. To apply both positive and negative weights, two consecutive intensity measurements are performed: the first applying only the positive weights, and the second applying only the (absolute values of the) negative weights. \textcolor{black}{We define $y^+$ and $y^-$ as intensity measurements obtained with positive and negative weights, respectively.} The output signal $y$ is generated as their linear combination:
\begin{equation}
\label{eq:optical_output}
    y = C^+\cdot y^+ + C^-\cdot y^- + C^0,
\end{equation}
where $(C^+,\,C^-,\,C^0)$ is a set of coefficients optimized for each task. The operation in Eq.\ \eqref{eq:optical_output} is currently executed on a digital computer, but it can be implemented fully in the analog domain. This same scheme for optical weighting has been presented in \cite{Lupo2021} and \cite{Butschek2022}. We also reported the design for a fully integrated optical output layer based on a similar principle in Refs.~\cite{Jonuzi2023A, Jonuzi2023}.

\subsection{\textcolor{black}{Experimental Evaluation.}}\label{sec:experimental_verif}
The evaluation of the system performance has been conducted as follows. First, we tested our photonic ELM on three publicly available classification benchmark tasks: the Heart disease\cite{heart_dataset}, the Wine\cite{wine_dataset}, and the Iris\cite{iris_dataset} classification task. In this phase, 
we compared many possible operating conditions of the system (\textcolor{black}{i.e. encoding the input in comb-line intensities vs. comb line phases or mixing the information through Kerr effect with and without additional phase modulation}). 

After having identified the best-performing configuration \textcolor{black}{using classification datasets}, we \textcolor{black}{transfer the system parameters to test its performance on a regression task.} The \textcolor{black}{selected benchmark predicts real-estate values}, based on the dataset first presented in \textcolor{black}{Ref.}~\cite{Yeh2018}. \textcolor{black}{For every tested configuration, we evaluated the system performance for both, digital and optical weighting. For both types of weighting, we tested the system with and without an optional propagation in a 540-meter-long SMF.} 

We also compared the performance of the photonic ELM with a software ELM counterpart \textcolor{black}{with} 25 hidden neurons and a quadratic or quartic nonlinearity. The quadratic nonlinearity is inspired by our previous work \cite{Lupo2021}, where the only existing nonlinearity is due to the output-layer photodiode. The quartic nonlinearity is inspired by \textcolor{black}{our model for fiber-output comb-line intensities} (Eq.\ \eqref{eq:KerrEffect} and App.~\ref{appendix:A}). In the case of the Wine and Iris datasets, we compared the performance of the current implementation with the results experimentally obtained in the previously reported scheme that uses only phase modulation for information mixing \cite{Lupo2021}. 

For every benchmark, 70\% of each dataset \textcolor{black}{were} used for training the output weights and 30\% for testing the predictions. The classification performance \textcolor{black}{was} evaluated in terms of accuracy, i.e.\ the percentage of correctly classified examples over their total number. \textcolor{black}{The performance on the regression task was evaluated by means of the root-mean-square error (RMSE):}
\begin{equation}
\label{eq:rmse}
\textit{RMSE}=\sqrt{\frac{1}{N}\sum_1^N\left(y_i-\hat{y}_i\right)^2},
\end{equation}
where $N$ is the number of regressions made, $y_i$ is the predicted value for the $i$-th regression and $\hat{y}_i$ is the expected value for the $i$-th regression. For each benchmark test, we repeated the training procedure 100 times selecting different partitions in train and test datasets. We recorded both\textcolor{black}{,} the average and standard deviation of the accuracy values.

\section{Results}
\label{sec:results}

In this section\textcolor{black}{,} we report the experimental results measured over multiple configurations of the system, as described in Sec.\ \ref{sec:experimental_verif}. To distinguish among the possible configurations, we employ the expressions ``amplitude'' and ``phase'' to specify the encoding mechanism (whether data is encoded in comb line amplitudes or phases, respectively). \textcolor{black}{Also, we use} the expressions ``EDFA only'', ``EDFA+SMF'' and ``EDFA+SMF+PM'' to specify the configuration of the setup (cf. Fig.\ \ref{fig:ExpSetup}).

The first set of measurements \textcolor{black}{is focussed on three classification benchmarks and serves} the purpose of identifying the best encoding method (amplitude \textcolor{black}{vs.} phase) \textcolor{black}{as well as} the best experimental configuration (``EDFA+SMF'' \textcolor{black}{vs.} ``EDFA+SMF+PM''). Encoding in amplitude in the ``EDFA+SMF' configuration clearly emerged as the best performing option (we discuss this in the next paragraphs). For ease of readability, the results of this first set of measurements are reported in App.~\ref{appendix:B}, Tab.~ \ref{tab:results1}. 

\begin{figure}
\centering
\begin{subfigure}[b]{0.83\textwidth}
    \includegraphics[width=\linewidth]{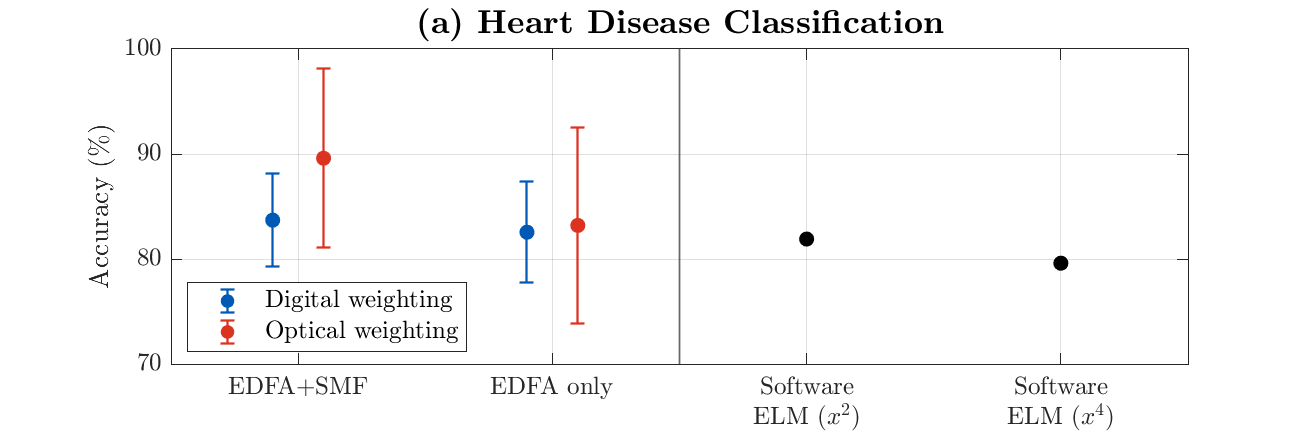}    
\end{subfigure}

\vspace{15 px}

\begin{subfigure}[b]{0.83\textwidth}
    \includegraphics[width=\linewidth]{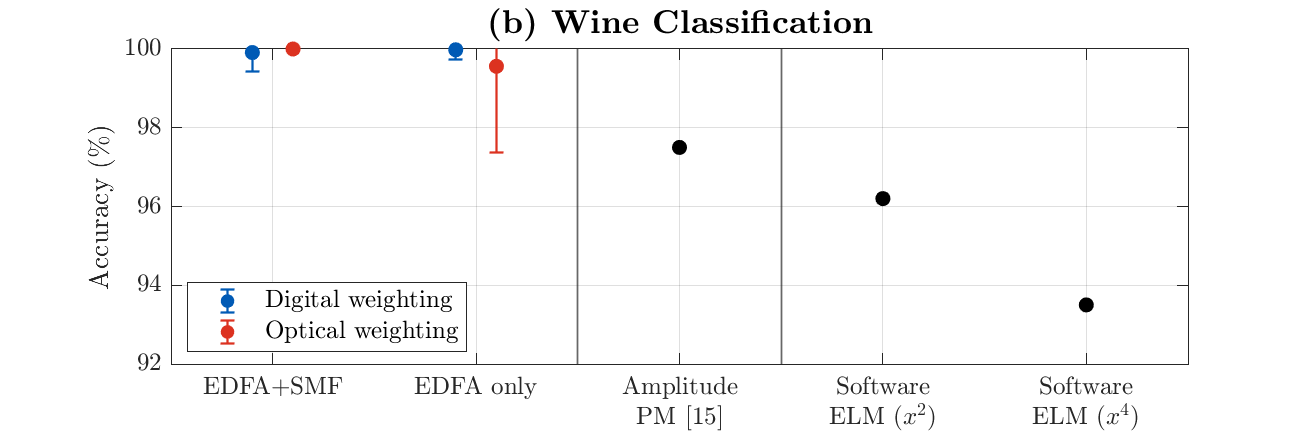}  
\end{subfigure}

\vspace{15 px}

\begin{subfigure}[b]{0.83\textwidth}
    \includegraphics[width=\linewidth]{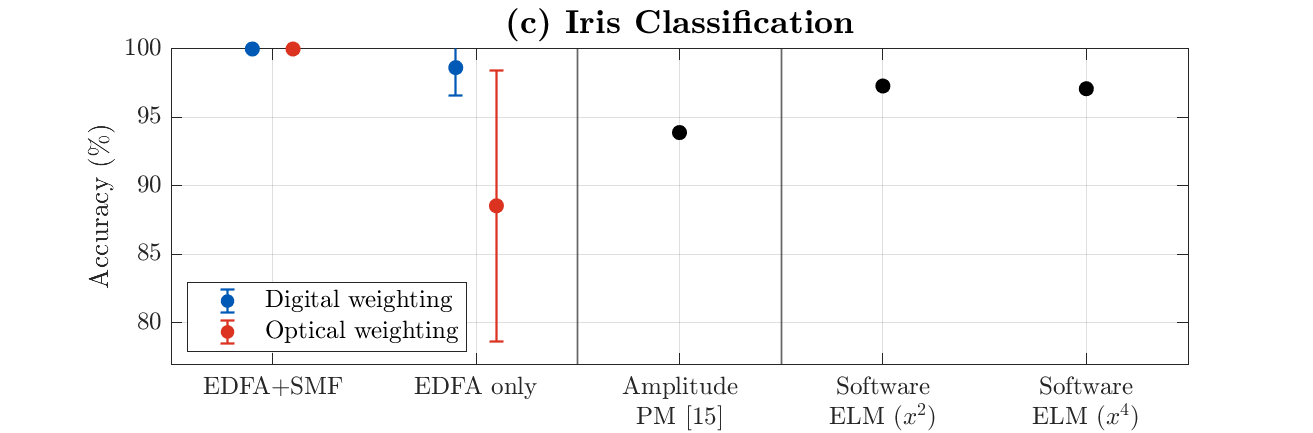}    
\end{subfigure}

\vspace{15 px}

\begin{subfigure}[b]{0.83\textwidth}
    \includegraphics[width=\linewidth]{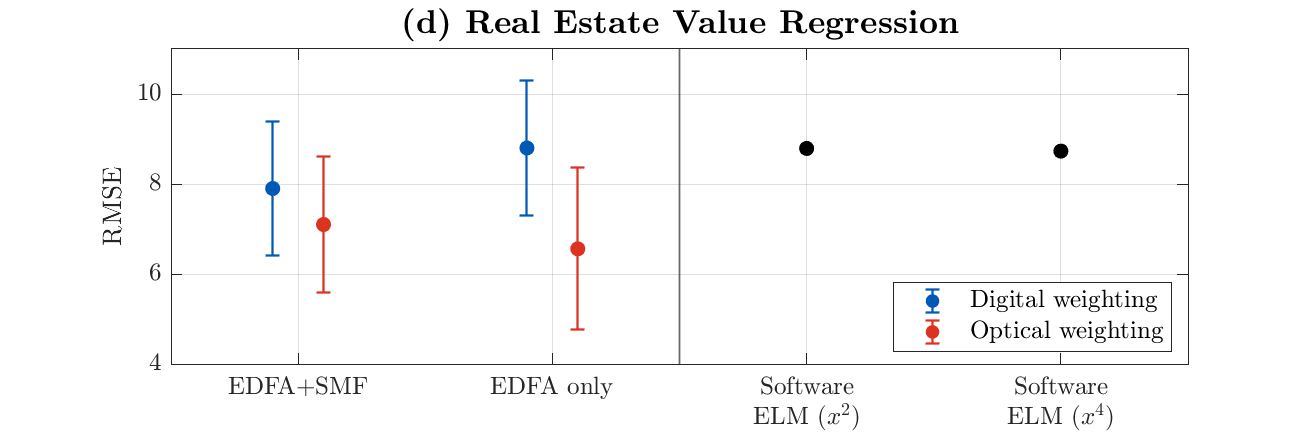}    
\end{subfigure}

\caption{Experimental results obtained by employing \textcolor{black}{digital (blue) and optical (red) weighting, encoding the input in amplitudes, and using "EDFA+SMF" and "EDFA only" configurations. They represent average scores achieved} by randomly selecting the train and test subsets 100 times. The error bars represent the standard deviation of the accuracy distribution. \textcolor{black}{Experimental results are compared to} fairly equivalent software ELMs (25 hidden nodes, quadratic or quartic nonlinearity) and to a previous experiment based on information mixing via phase modulation only \cite{Lupo2021}.}
\label{fig:results2}
\end{figure}

\textcolor{black}{The} second set of measurements \textcolor{black}{was} executed to test in depth the system in the best configuration (encoding in amplitude, ``EDFA+SMF'').
Initially, we thought that the EDFA did not play any role in the information mixing \textcolor{black}{process, but only helped} boost the power so that \textcolor{black}{the actual} information mixing could take place in the \textcolor{black}{540-long SMF}. However\textcolor{black}{, a} careful analysis of our data showed that significant information mixing \textcolor{black}{took} place in the EDFA itself. \textcolor{black}{We, thus,} extended the study to include the ``EDFA only' configuration.
This second \textcolor{black}{measurement set included both,} the classification and the regression benchmarks. \textcolor{black}{The achieved results} are reported in Fig.~\ref{fig:results2}. In this figure, we also compare \textcolor{black}{the} results with what was obtained from the previous experiment (\textcolor{black}{in which the information mixing was performed by a} PM only \cite{Lupo2021}). \textcolor{black}{We also compare them with} the results from \textcolor{black}{an} equivalent software-based ELMs (\textcolor{black}{with} 25 hidden neurons \textcolor{black}{and a} quadratic or quartic nonlinearity). The results are summarized in App.~\ref{appendix:B}, Tabs.~\ref{tab:results2} and \ref{tab:results3}.

%
%
%
%
%
%

The recorded performance depends \textcolor{black}{on both, the input-data encoding scheme (amplitude vs. phase) and the mixing technique (i.e. the experimental configuration)}. \textcolor{black}{The encoding} scheme and mixing technique \textcolor{black}{can} be considered as hyperparameters of the network and, as usual in machine learning, the best-performing hyperparameters depend on the task. Nevertheless, we can derive two general observations from the results. 
First, \textcolor{black}{the} encoding in amplitude generally performs better than encoding in phase. \textcolor{black}{This result supports our model that shows that the information mixing works via the comb-line intensities rather than their phases and is driven by FWM (Eq.~\ref{eq:KerrEffect}).} 
Second, \textcolor{black}{an} additional phase modulation \textcolor{black}{by PM2} is never useful and in some cases even detrimental. Phase modulation generates \textcolor{black}{only linear mixing between the neurons} (Eq.~\eqref{eq:pm}).
Apparently, the nonlinear propagation in the fiber already introduces sufficient data mixing, thus additional linear mixing provided by the phase modulation does not improve the performance.
Apart from that, the electro-optical phase modulator has a polarization-dependent loss (nominally 5 dB) that might introduce fluctuations of the output power since thermal effects can rotate the polarization state. These fluctuations, when happening during the training phase, might result in a (further) reduction of the performance.

In digital weighting mode, the photonic ELM achieved accuracies of $83.7\%\pm4.4\%$, $99.9\%\pm0.5\%$, and $100\%\pm0\%$ on the Heart diseases, Wine, and Iris classification benchmarks respectively. In optical weighting mode, the photonic ELM achieved accuracies of $89.6\%\pm8.5\%$, $100\%\pm0\%$, and $100\%\pm0\%$ on the Heart diseases, Wine, and Iris classification benchmarks respectively. Concerning the regression, the photonic ELM achieved an RMSE of $8.52\pm1.65$ and $8.18\pm2.06$ in digital and optical weighting modes respectively. These results refer to the best-performing configuration for each task, on the test fraction of the dataset.

In the best-performing configuration, optical weighing always delivers results comparable to digital weighting. In some cases, optical weighting average accuracies are slightly higher than \textcolor{black}{digital-weighting} ones. This is probably due to \textcolor{black}{some} statistical fluctuations. However, we note \textcolor{black}{that, in optical-weighting mode,} the positive-weighted signal\textcolor{black}{s} and the negative-weighted signal\textcolor{black}{s} are combined through optimized coefficients (Eq.~\eqref{eq:optical_output}). In principle, this combination should be linear. \textcolor{black}{In practice, though,} an extra nonlinearity may be introduced by the saturation of \textcolor{black}{the} photodiodes\textcolor{black}{, which might explain} the increased performance.

When compared \textcolor{black}{to} our previous experiment (Fig.\ \ref{fig:results2} and App.~\ref{appendix:B}, Tab.~\ref{tab:results3}), the approach based on \textcolor{black}{the amplitude-encoded data mixed by the Kerr nonlinearity} delivers better performance in both (digital and optical) weighting modes.

\textcolor{black}{Our experimental scheme also slightly outperformed the software-based ELM counterparts that were designed to replicate the experimental information mixing (with quadratic or quartic nonlinearity). This indicates that the experimental ELM might benefit from nonlinearities that are stronger and/or richer than the quadratic and quartic ones. This could be due to higher-order contributions of the Kerr effect that are not captured by the quartic approximation, as well as nonlinearities in other experimental devices (e.g., the saturation of the photodiodes).} 

\begin{figure}
\centering
\begin{subfigure}[b]{0.83\textwidth}
    \includegraphics[width=\linewidth]{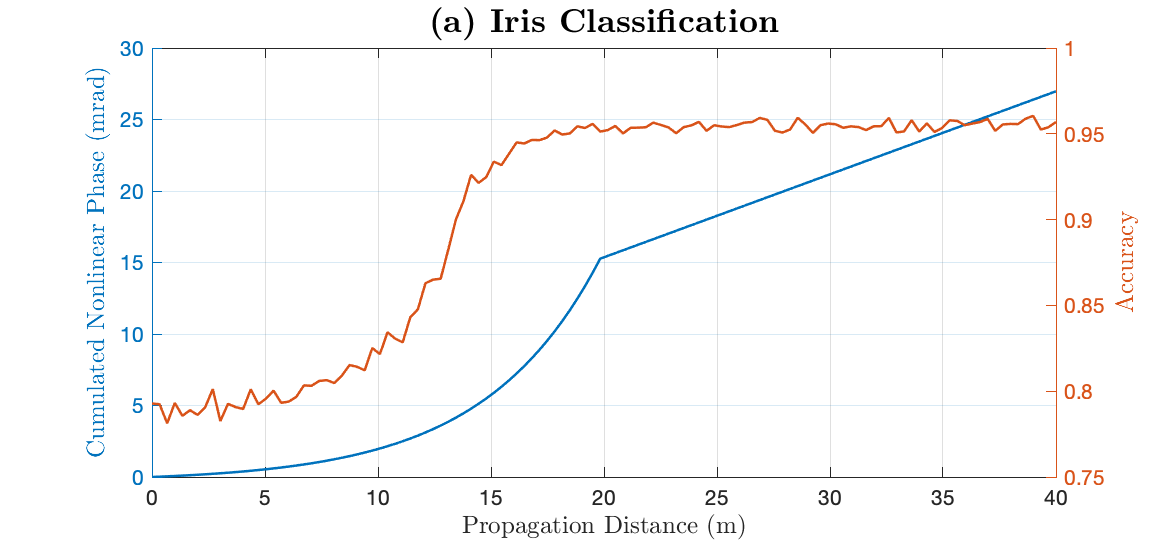}    
\end{subfigure}

\vspace{15 px}

\begin{subfigure}[b]{0.83\textwidth}
    \includegraphics[width=\linewidth]{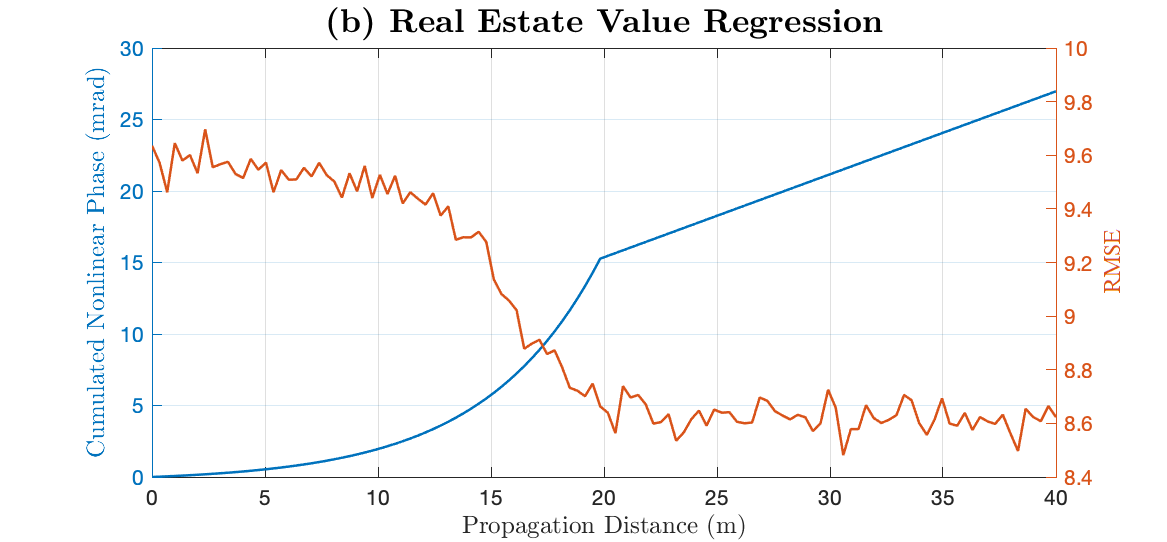}  
\end{subfigure}

\caption{Results of the numerical simulation based on Eq.\ \eqref{eq:GLNS} for two benchmark tasks: Iris \textcolor{black}{c}lassification (a) and Real Estate Value \textcolor{black}{r}egression (b). The initial $20$ meters of propagation take place in an \textcolor{black}{e}rbium-doped (amplifying) fiber, while the last $20$ meters happen in a standard telecom fiber (Sec.\ \ref{sec:setup}). The cumulated nonlinear phase (left axis) is given by $\Phi(L)=\int_0^L\gamma(z)P(z)dz$ where the $z$ dependence of $\gamma$ takes into account that the nonlinear coefficient is different in the EDFA and in the standard fiber, while the $z$ dependence of $P(z)$ takes into account amplification and loss.
The discontinuity of $\Phi(L)$  at $L=20$ m is because the \textcolor{black}{e}rbium-doped fiber is amplifying, thus, $P(z)$ exponentially increases with length, while the standard telecom fiber is not amplifying, thus, $P(z)$ slightly decreases due to attenuation. The results of this figure were obtained using the 4th-order Runge-Kutta in the Interaction Picture Method (RK4IP) including a realistic description of the erbium-doped fiber \cite{Balac16}.}\label{fig:simulation}
\end{figure}

\textcolor{black}{Further, we} observed that propagation \textcolor{black}{only} in the erbium-doped fiber is enough to perform information processing, although the addition of the 540-meter-long SMF systematically results in \textcolor{black}{a} slightly increased performance. 
To better clarify the effect of the propagation, we numerically integrate Eq.\ \eqref{eq:GLNS} to evaluate both the performance and accumulated nonlinear phase as a function of propagation length.  \textcolor{black}{The results} are reported in Fig.\ \ref{fig:simulation}. \textcolor{black}{They show} that the propagation in the EDFA ($L<20$ m) is sufficient for the ELM performance to approach the best value. \textcolor{black}{We reduced the plots} to a length of $40$ meters because this is the most interesting region\textcolor{black}{,} no performance increase was detected for longer propagation \textcolor{black}{distances.}

\section{Conclusion}
\label{sec:conclusion}
We presented a photonic implementation of a frequency-multiplexed photonic Extreme Learning Machine \textcolor{black}{that encodes input information in the lines of a frequency comb and exploits the Kerr nonlinearity in optical fibers to process it. In our system, the information processing occurs via Kerr-driven four-wave mixing, rather than via self- or cross-phase modulation or dispersion effects, which is shown by a perturbative model we derived and presented here.} We experimentally validated \textcolor{black}{the Extreme Learning Machine using} four different benchmark tasks \textcolor{black}{including classification and regression problems.}
\textcolor{black}{Optimizing the system performance, we} compared multiple possible configurations, varying the input encoding method (\textcolor{black}{amplitude vs. phase), way of information mixing (Kerr nonlinearity with and without a subsequent phase modulation), and the kind of output weights (digital vs. optical).} The best results were \textcolor{black}{achieved} by encoding \textcolor{black}{the} neurons in \textcolor{black}{amplitudes of the frequency-comb lines and using only the Kerr for information mixing.} Digital and optical weighting \textcolor{black}{provided} comparable results. Encoding the neurons in the phases \textcolor{black}{of the comb lines resulted in a worsening of the EML performance}.

Surprisingly, our work shows that a weak Kerr nonlinearity can yield good results on classification and regression tasks. In fact the propagation in the $20$-meter-long erbium-doped fiber included in the amplification stage is already sufficient to process information. The amount of nonlinearity can be quantified by the product $\phi = \gamma L P$, which in our work is either $\phi \approx 0.0015$ rad or $\phi \approx 0.3$ rad
 depending on the configuration, corresponding to a very weak nonlinearity. All our results, theoretical, experimental, numerical-simulations, coincide well with each other showing that a weak Kerr nonlinearity is a highly efficient mechanism for information mixing in our frequency-multiplexed fiber-based Extreme Learning Machine.

\textcolor{black}{Let us now briefly compare our Extreme Learning Machine with approaches presented in Refs.~\cite{Zhou2022} and ~\cite{Fischer2023}. In these references, the authors use broad continuous spectra from femtosecond pulsed lasers as sources to modulate the input data onto them by deploying programmable spectral filters. Then, they propagate modulated spectra in pieces of highly nonlinear fibers to allow for data transformation and collect fiber outputs for further evaluation on a PC. Both papers point out the importance of the nonlinear dynamics described by the Nonlinear Schr\"odinger equation. Intuitively, they expect that higher values of optical Kerr nonlinearity would lead to a better performance. This is why their setups operate in highly nonlinear regimes
(for instance in Ref. \cite{Zhou2022}  the nonlinear phase is $\phi \approx 165$ rad).
 This intuition is justified as theoretical studies from Ref.~\cite{Marcucci2020} suggest. Contrary to it, we show that already a weak Kerr nonlinearity can be used for effective data processing.} 

\textcolor{black}{
As the information-carrying spectra from Refs.~\cite{Zhou2022} and ~\cite{Fischer2023} are continuos, they are subject to several effects such as, for instance, dispersion, self-phase modulation, intrapulse stimulated Raman scattering, and four-wave mixing. All these effects occur simultaneously and impact the dynamics in a complex, delicate, and interconnected manner. In our frequency-multiplexed Extreme Learning Machine operating with a discrete spectrum under weak Kerr nonlinearity, the only really decisive effect is four-wave mixing, while dispersion plays a minor role. In general, the dynamics of systems that operate in highly nonlinear regime are sensitive to the input noise. This dependence can severely affect the data-processing capability of a neuromorphic scheme. From this perspective, weak-nonlinearity schemes like ours can offer the possibility of more robust neuromorphic-computing implementations.  One way or another, both approaches (with strong or weak Kerr nonlinearity) have advantages and disadvantages and should be pursued in future. For both approaches, it is important to understand the evolution of the data-encoded optical signal and how various effects might enhance or impact the overall data-processing capability of the neuromorphic scheme.}

\textcolor{black}{
In future, more complex task with more input features should be addressed. For this, we would need a broader frequency comb to encode the data. A broader comb can be generated using the beating between two continuous-wave lasers, rather than deploying a phase modulator at the initial stage (cf.~\cite{zajnulina_2015OFC}). To increase the data throughput, we will replace the quite slow programmable spectral filters we currently deploy by photonic-integration designs we present in Refs.~\cite{Jonuzi2023A}, \cite{Jonuzi2023}. In fact, our whole Extreme Learning Machine can be transferred to a chip. Semiconductor-based solutions needed for on-chip integration are easily available (e.g. semiconductor optical amplifiers incl. erbium-doped waveguides \cite{Liu2022} as well as frequency-comb sources \cite{zajnulina_soa2017}, \cite{Gaeta2019}). Photonic integration of our Extreme Learning Machine is attractive as it will allow for lowering the power consumption while contributing to scalability, stability, and practicability.}   

\section*{Acknowledgement} 
We acknowledge interesting and inspiring conversations with Sonia Boscolo (Aston University Birmingham, UK), Christophe Finot (ICB - Université de Bourgogne, France), and Ingo Fischer (IFISC - Universitat de les Illes Balears, Spain).

\section*{Funding}
We acknowledge the funding from H2020 Marie Skłodowska-Curie Actions (Project POST-DIGITAL, grant number 860830), FWO and F.R.S.-FNRS Excellence of Science (EOS) program grant 40007536, \textcolor{black}{and Win4Space project of Walloon's Region excellence programme}.

\section*{Disclosures}
The authors declare no conflicts of interest.

\appendix

\section{Perturbative approximation of the Kerr effect on comb line intensities} \label{appendix:A}

\subsection{\textcolor{black}{Nonlinear Schr\"odinger Equation}}

The propagation along the fiber is described by the Nonlinear Schr\"odinger Equation (NLS) \cite{Agrawal}:
\begin{equation}
\partial_z x(z,t) =  i \beta(i\partial_t)x(z,t) - \frac{\alpha}{2}x(z,t) - i \gamma \vert x(z,t) \vert^2 x(z,t),
\label{Eq:NLS}
\end{equation}
where $x(z,t)$ is the electric field, $\beta(\omega)$ the propagation constant, $\alpha$ the attenuation (negative $\alpha$ corresponds to the gain), and $\gamma$ the nonlinear Kerr coefficient of the fiber. 

\textcolor{black}{The total power at the propagation point $z$ is given by} $P(z)=\int dt \vert x(z,t)\vert^2.$ If $\beta(i\partial_t)$ is a Hermitian operator, then the total power satisfies \textcolor{black}{the equation}$ \frac{d P(z)}{dz} = -\alpha P(z)$. \textcolor{black}{Therefore, we have:}
\begin{equation}
P(z)=P_0 e^{-\alpha z}
\label{Eq:P(z)}
\end{equation}
with $P_0$ \textcolor{black}{being} the initial total power.

According to Sec.\ \ref{sec:principles}, the input field, thus, the initial condition, is a frequency comb:
\begin{equation}
x(z=0,t) = \sum_k x_k e^{-i \Omega k t},
\end{equation}
where $x_k$ are the (complex) amplitudes of the input comb lines and $\Omega$ is the spacing \textcolor{black}{between the} comb lines.

If $\gamma = 0$, then the solution \textcolor{black}{of the NLS reads as:} 
\begin{equation}
x(z,t) = \sqrt{P_0} \sum_k b_k^0 e^{-i \Omega k t} e^{ i \beta(\Omega k)z} e^{ - \frac{\alpha}{2}z},
\end{equation}
\textcolor{black}{which shows that} each comb line acquires a different phase due to \textcolor{black}{the} dispersion ($\beta(\Omega k)z$) and all comb lines are attenuated identically ($- \frac{\alpha}{2}z$). 

If $\gamma \neq 0$, we write the solution as:
\begin{eqnarray}
x(z,t) &=& \sum_k x_k(z) e^{-i \Omega k t}\nonumber\\
&=& \sqrt{P_0} e^{ - \frac{\alpha}{2}z} \sum_k b_k(z) e^{-i \Omega k t} e^{ i \beta(\Omega k)z}, 
\label{Eq:Ansatz}
\end{eqnarray}
\textcolor{black}{which shows} that the \textcolor{black}{temporal} comb shape is preserved. \textcolor{black}{However,} the amplitudes and phases of the comb lines acquire a  $z-$dependence due to the Kerr nonlinearity. \textcolor{black}{Here, $b_k(z)$ are normalized comb-line amplitudes, i.e.
\begin{equation}
\sum_k \vert b_k(z) \vert^2 = 1,\ 
\label{Eq:normb_k}
\end{equation}
to satisfy Eq.~\eqref{Eq:P(z)}}. 

Inserting Eq.\ \eqref{Eq:Ansatz} in Eq.\ \eqref{Eq:NLS}\textcolor{black}{, we obtain:}
\begin{equation}
\sum_k \partial_z b_k e^{-i \Omega k t} e^{ i \beta(\Omega k)z} e^{ - \frac{\alpha}{2}z}
= - i\gamma P_0 \sum_{l,m,n}
b_l \overline b_m b_n 
e^{-i \Omega (l-m+n) t}
e^{ i \left[ \beta(\Omega l) - \beta(\Omega m) + \beta(\Omega n) \right] z}
e^{ - \frac{3 \alpha}{2}z}.
\label{equ:comblines_all}
\end{equation}
\textcolor{black}{W}e can rewrite Eq.~\ref{equ:comblines_all} as equations for each comb line amplitude:
\begin{equation}
 \partial_z b_k 
= - i\gamma P_0 e^{ -  \alpha z} \sum_{ l-m+n=k}
b_l \overline b_m b_n 
e^{ i \left[ \beta(\Omega l) - \beta(\Omega m) + \beta(\Omega n)- \beta(\Omega k)  \right] z}
\ .
\label{Eq:NLS2}
\end{equation}
\textcolor{black}{Fig.\ \ref{fig:Simul}(a) shows the comparison between the numerical integration of Eq.\ \eqref{Eq:NLS2} and the NLS integration via the split-step Fourier method (SSFM). The results deviate only slightly, which is probably due to a limited numerical precision. Thus, we conclude that Eq.~\ref{Eq:NLS2} represents a good description of the comb-line evolution in optical fibers.}

\subsection{Self- and Cross-Phase Modulation}
The $z-$dependent phase on the right-hand side of Eq.\ \eqref{Eq:NLS2} vanishes if \textcolor{black}{the sum indices}
$l$ or $n$ equal $k$. These terms correspond to \textcolor{black}{self- and cross-phase modulation.} If we restrict the right-hand side of Eq.~\eqref{Eq:NLS2}  to the terms such that $l$ or $n$ equal $k,$ we have the equations:
\begin{eqnarray}
 \partial_z b_k 
&=& - i\gamma  P_0 e^{ -  \alpha z}  \left(2\sum_{ l\neq k} \vert b_l\vert^2  + \vert b_k\vert^2 \right)  b_k
e^{ -  \alpha z}\nonumber\\
&=& -i \gamma P_0 e^{ -  \alpha z}  (2- \vert b_k\vert^2 ) b_k,
\label{Eq:NLS-XSPM}
\end{eqnarray}
where we used Eq. \eqref{Eq:normb_k}. Eq.\ \eqref{Eq:NLS-XSPM} is \textcolor{black}{solved by:}
\begin{equation}\label{eq:solutionSPM}
b_k(z) = e^{-i \phi_k(z)} b_k^0,
\end{equation}
with\begin{equation}\label{eq:phi}
\phi_k(z) = \gamma P_0  (2- \vert b_k^0\vert^2 )  \frac{1 - e^{-\alpha z} }{\alpha}\ .
\end{equation}
Here, $b_k^0$ is the initial normalized \textcolor{black}{comb-line} amplitude at $z=0$. Eq.~\eqref{eq:solutionSPM} \textcolor{black}{shows} that only the phase of $b_k$ is affected by \textcolor{black}{self- and cross-phase modulation, not its amplitude.} \textcolor{black}{In our implementation of a photonic ELM, self- and cross-phase modulation have no measurable effect as we read out only the comb intensities.}

\subsection{\textcolor{black}{Four-Wave Mixing}}

Eq.~\eqref{Eq:NLS-XSPM} neglects \textcolor{black}{FWM} terms in Eq.\ \eqref{Eq:NLS2}. To take them into \textcolor{black}{account, we carry out a perturbative expansion in $\gamma$. For this, we perturb the comb-line amplitudes by a small function $\epsilon_k(z):$}
\begin{equation} 
x(z,t) = \sqrt{P_0} \sum_k  e^{-i \Omega k t} e^{ i \beta(\Omega k)z} e^{-i \phi_k(z)} e^{ - \frac{\alpha}{2}z} 
 \left(  b_k^0 + \gamma \epsilon_k(z)\right),
\label{Eq:gammaEps}
\end{equation}
where $\phi_k(z)$ given by Eq.\ \eqref{eq:phi} and takes into account the \textcolor{black}{self- and cross-phase} modulation.

Upon inserting Eq.\ \eqref{Eq:gammaEps} into the NLS (Eq.\ \eqref{Eq:NLS}), the terms of order $\gamma^0$ cancel, while the terms of order $\gamma^1$  give an equation for $\epsilon_k$(z):
\begin{equation}
 \partial_z \epsilon_k
=  -i P_0 e^{ - \alpha z}
\sum_{l-m+n=k; n\neq k ; l\neq k}
b_l^0 \overline {b_m^0} b_{n}^0  
e^{ i \Delta \beta_{k,l,m,n}  z} + O(\gamma)
\label{Eq:eqepsilonk}
\end{equation}
where the condition on the summation indices $l$, $m$, and $n$ expresses that \textcolor{black}{self- and cross-phase} terms are excluded, and we denote
\begin{equation}
\Delta \beta_{k,l,m,n}  = \beta(\Omega l) - \beta(\Omega m) +
\beta(\Omega n) - \beta(\Omega k) \ .
\label{eq:deltabeta}
\end{equation}
Note that we have omitted the contribution of the phases $\phi_k(z)$ in Eq. \eqref{Eq:eqepsilonk} as these are of order $\gamma$.

The solution of Eq. \eqref{Eq:eqepsilonk} \textcolor{black}{gives us the perturbation function $\epsilon_k(z):$}
\begin{equation}
\epsilon_k (z)
= P_0 \sum_{l-m+n=k ; n\neq k ; l\neq k}
\frac{ b_l^0 \overline {b_m^0} b_{n}^0   }
{
 \Delta \beta_{k,l,m,n}  + i \alpha
} \left( 1 -
e^{ i\Delta \beta_{k,l,m,n} z}e^{ - \alpha z} \right).
\end{equation}
\textcolor{black}{It will allow us to determine the intensities of the comb lines in the next step.}

\subsection{\textcolor{black}{Comb-line Intensities}}
The intensity of \textcolor{black}{a comb-line} $k$ is obtained by taking the norm square of Eq.\ \eqref{Eq:gammaEps} to yield:
\begin{equation}
I_k = \vert x_k(z) \vert^2  =
P_0 e^{ - \alpha z}  \left( 
\vert b_k^0 \vert^2 \nonumber\\
+ \gamma  ( \overline {b_k^0} \epsilon_k (z) + c.c.)
+ O(\gamma^2) \right)\ .
\label{Eq:Ik}
\end{equation}

The change in \textcolor{black}{the intensity of this line} to order $\gamma^1$ (noted $\Delta I_k^1$), normalized by the total power $P(z)=P_0 e^{ - \alpha z}$ is
\begin{eqnarray}
\frac{ \Delta I_k^1}{P(z) } &=& 
\gamma   (\overline {b_k^0}  \epsilon_k  + c.c.) \nonumber\\
&=& \gamma P_0 \ Real\left[
\sum_{l-m+n=k ; n\neq k ; l\neq k}
\frac{ b_l^0 \overline {b_m^0} b_{n}^0 \overline {b_k^0}  }
{
 \Delta \beta_{k,l,m,n}  + i \alpha
} \left( 1 -
e^{ i\Delta \beta_{k,l,m,n} z}e^{ - \alpha z} \right) \right].
 \label{Eq:DeltaIk}
\end{eqnarray}
A comparison between the linearized change in intensity (Eq.\ \eqref{Eq:DeltaIk}) and the result of the numerical integration of NLS is given in Fig.\ \ref{fig:Simul}(b). With increasing propagation length, we see a deviation between the solution provided by NLS and Eq.~\eqref{Eq:DeltaIk}. 

Eq. \eqref{Eq:DeltaIk}  simplifies in the limit of small propagation lengths when
$e^{ i\Delta \beta_{k,l,m,n} z}e^{ - \alpha z}  \approx 1 + (i\Delta \beta_{k,l,m,n}  - \alpha )z,$ whereupon one finds:
\begin{equation}
\frac{ \Delta I_k^1}{P(z)} \approx - \gamma P_0 z \ Im
\left[
\sum_{l-m+n=k ; n\neq k ; l\neq k}
 b_l^0 \overline {b_m^0} b_{n}^0 \overline {b_k^0}  
 \right].
\end{equation}
\textcolor{black}{Here, we see that the nonlinearity that is introduced to the ELM network by the Kerr effect is of the fourth order, i.e. quartic.}

Eq. \eqref{Eq:DeltaIk} also simplifies in the case of strong amplification when $e^{ - \alpha z}  \gg 1$ (corresponding to propagation in the EDFA). Assuming also that $\alpha \gg \Delta \beta$, one finds
\begin{equation}
\frac{ \Delta I_k^1}{P(z)} \approx \gamma P_0  e^{ - \alpha z}  \ Im
\left[
\sum_{l-m+n=k ; n\neq k ; l\neq k}
 b_l^0 \overline {b_m^0} b_{n}^0 \overline {b_k^0}  e^{ i\Delta \beta_{k,l,m,n} z}
 \right].
\end{equation}

\section{\textcolor{black}{Summary of the Results}}
\label{appendix:B}
Here we present all our results in three tables.

\begin{table}[]
\centering
\begin{tabular}{cc|cc|cc|}
\cline{3-6} &                                                      & \multicolumn{2}{c|}{\textbf{Amplitude Encoding}}& \multicolumn{2}{c|}{\textbf{Phase Encoding}} \\ \hline
\multicolumn{1}{|c|}{\textbf{Task}}                                 & \textbf{\begin{tabular}[c]{@{}c@{}}Weighting\end{tabular}} & \multicolumn{1}{c|}{\textbf{EDFA+SMF}} & \textbf{\begin{tabular}[c]{@{}c@{}}EDFA+SMF+PM \end{tabular}} & \multicolumn{1}{c|}{\textbf{EDFA+SMF}} & \textbf{\begin{tabular}[c]{@{}c@{}}EDFA+SMF+PM\end{tabular}} \\ \hline
\multicolumn{1}{|c|}{\multirow{2}{*}{\begin{tabular}[c]{@{}c@{}}Heart\\ disease\end{tabular}}} & Digital                  & \multicolumn{1}{c|}{$83.7\%$} & $80.5\%$                              & \multicolumn{1}{c|}{$71.5\%$} & $60.6\%$                            \\ \cline{2-6} 
\multicolumn{1}{|c|}{}                                             & Optical                                                           & \multicolumn{1}{c|}{$89.6\%$} & $83.3\%$                             & \multicolumn{1}{c|}{$80.4\%$} & $56.1\%$                             \\ \hline
\multicolumn{1}{|c|}{\multirow{2}{*}{Wine}}                         & Digital                                                           & \multicolumn{1}{c|}{$99.9\%$}     & $99.3\%$                          & \multicolumn{1}{c|}{$92.8\%$}     & $96.7\%$                          \\ \cline{2-6} 
\multicolumn{1}{|c|}{}                                              & Optical                                                           & \multicolumn{1}{c|}{$100\%$}     & $91.5\%$                          & \multicolumn{1}{c|}{$95.4\%$}     & $78.0\%$                          \\ \hline
\multicolumn{1}{|c|}{\multirow{2}{*}{Iris}}                         & Digital                                                           & \multicolumn{1}{c|}{$100\%$}      & $100\%$                           & \multicolumn{1}{c|}{$100\%$}      & $99.8\%$                          \\ \cline{2-6} 
\multicolumn{1}{|c|}{}                                              & Optical                                                           & \multicolumn{1}{c|}{$100\%$}    & $99.8\%$                         & \multicolumn{1}{c|}{$78.2\%$}    & $87.7\%$                            \\ \hline
\end{tabular}
\caption{Results for the first measurement set. Three classification tasks were used in \textcolor{black}{four} experimental schemes. From these \textcolor{black}{results, it appears that the amplitude encoding of input information is  better than the phase encoding. Also, we see that using a phase modulator for information mixing does not improve the performance of the ELM.} 
}\label{tab:results1}
\end{table}

\begin{table}[]
\centering
\begin{tabular}{|c|c|c|c|}
\hline
\multirow{2}{*}{\textbf{Task}} & \multirow{2}{*}{\textbf{Weighting}} & \textbf{EDFA+SMF}    & \textbf{EDFA only}   \\
                               &                                     & (Amplitude encoding) & (Amplitude encoding) \\ \hline
Heart disease                  & Digital                             & 83.7\%               & 82.6\%               \\ \cline{2-4} 
(Accuracy)                     & Optical                             & 89.6\%               & 83.2\%               \\ \hline
Wine                           & Digital                             & 99.9\%               & 100\%                \\ \cline{2-4} 
(Accuracy)                     & Optical                             & 100\%                & 99.6\%               \\ \hline
Iris                           & Digital                             & 100\%                & 98.6\%               \\ \cline{2-4} 
(Accuracy)                     & Optical                             & 100\%                & 88.6\%               \\ \hline
Real estate                    & Digital                             & 8.52                 & 8.91                 \\ \cline{2-4} 
(RMSE)                         & Optical                             & 8.96                 & 8.18                 \\ \hline
\end{tabular}
\caption{Results for the second measurement set. \textcolor{black}{All benchmarks were executed for the best performing configuration selected after the first measurement set (amplitude encoding, without phase modulation). Each task was evaluated with and without the propagation in a 540-meter-long single-mode fiber. Adding an SMF after the EDFA generally results in a small performance improvement.} 
}\label{tab:results2}
\end{table}

\begin{table}[]
\centering
\begin{tabular}{|c|cc|c|}
\hline
\multirow{2}{*}{\textbf{Task}}                                     & \multicolumn{2}{c|}{\textbf{Software ELM}}     & \multicolumn{1}{l|}{\textbf{Previous experiment}} \\ \cline{2-3}
                                                                   & \multicolumn{1}{c|}{Quadratic NL} & Quartic NL & \multicolumn{1}{l|}{(Amplitude, PM only)}         \\ \hline
\begin{tabular}[c]{@{}c@{}}Heart disease\\ (Accuracy)\end{tabular} & \multicolumn{1}{c|}{81.9\%}       & 79.6\%     & Not measured                                      \\ \hline
\begin{tabular}[c]{@{}c@{}}Wine\\ (Accuracy)\end{tabular}          & \multicolumn{1}{c|}{97.5\%}       & 96.2\%     & 97.5\%                                            \\ \hline
\begin{tabular}[c]{@{}c@{}}Iris\\ (Accuracy)\end{tabular}          & \multicolumn{1}{c|}{97.3\%}       & 93.5\%     & 93.9\%                                            \\ \hline
\begin{tabular}[c]{@{}c@{}}Real estate\\ (RMSE)\end{tabular}   & \multicolumn{1}{c|}{8.79}             &    8.73        & Not measured                                      \\ \hline
\end{tabular}
\caption{Results obtained with a software ELM as well as experimental results from the previous work where information mixing was facilitated by a phase modulator \cite{Lupo2021}. }
\label{tab:results3}
\end{table}

\clearpage

\bibliography{bibl}

\begin{thebibliography}{10}
\newcommand{\enquote}[1]{``#1''}

\bibitem{Beyond2020}
\enquote{Beyond {v}on {N}eumann,} {\protect\JournalTitle{Nature
  Nanotechnology}} \textbf{15}, 507--507 (2020).

\bibitem{Reuther2019}
A.~Reuther, P.~Michaleas, M.~Jones, V.~Gadepally, S.~Samsi, and J.~Kepner,
  \enquote{Survey and benchmarking of machine learning accelerators,} in
  \emph{2019 {IEEE} High Performance Extreme Computing Conference ({HPEC}),}
  ({IEEE}, 2019).

\bibitem{Markovi2020}
D.~Markovi{\'{c}}, A.~Mizrahi, D.~Querlioz, and J.~Grollier, \enquote{Physics
  for neuromorphic computing,} {\protect\JournalTitle{Nature Reviews Physics}}
  \textbf{2}, 499--510 (2020).

\bibitem{Nakajima2020}
K.~Nakajima, \enquote{Physical reservoir computing{\textemdash}an introductory
  perspective,} {\protect\JournalTitle{Japanese Journal of Applied Physics}}
  \textbf{59}, 060501 (2020).

\bibitem{huang2006extreme}
G.-B. Huang, Q.-Y. Zhu, and C.-K. Siew, \enquote{Extreme learning machine:
  theory and applications,} {\protect\JournalTitle{Neurocomputing}}
  \textbf{70}, 489--501 (2006).

\bibitem{huang2015extreme}
G.-B. Huang, \enquote{What are extreme learning machines? filling the gap
  between frank rosenblatt’s dream and john von neumann’s puzzle,}
  {\protect\JournalTitle{Cognit. Comput.}} \textbf{7}, 263--278 (2015).

\bibitem{Huang2006}
G.-B. Huang, L.~Chen, and C.-K. Siew, \enquote{Universal approximation using
  incremental constructive feedforward networks with random hidden nodes,}
  {\protect\JournalTitle{{IEEE} Transactions on Neural Networks}} \textbf{17},
  879--892 (2006).

\bibitem{Marcucci2020}
G.~Marcucci, D.~Pierangeli, and C.~Conti, \enquote{Theory of neuromorphic
  computing by waves: Machine learning by rogue waves, dispersive shocks, and
  solitons,} {\protect\JournalTitle{Physical Review Letters}} \textbf{125}
  (2020).

\bibitem{Shastri2021}
B.~J. Shastri, A.~N. Tait, T.~F. de~Lima, W.~H.~P. Pernice, H.~Bhaskaran, C.~D.
  Wright, and P.~R. Prucnal, \enquote{Photonics for artificial intelligence and
  neuromorphic computing,} {\protect\JournalTitle{Nature Photonics}}
  \textbf{15}, 102--114 (2021).

\bibitem{saade2016}
A.~Saade, F.~Caltagirone, I.~Carron, L.~Daudet, A.~Dremeau, S.~Gigan, and
  F.~Krzakala, \enquote{Random projections through multiple optical scattering:
  Approximating kernels at the speed of light,} in \emph{2016 {IEEE}
  International Conference on Acoustics, Speech and Signal Processing
  ({ICASSP}),}  ({IEEE}, 2016).

\bibitem{Pierangeli21}
D.~Pierangeli, G.~Marcucci, and C.~Conti, \enquote{Photonic extreme learning
  machine by free-space optical propagation,} {\protect\JournalTitle{Photonics
  Res.}} \textbf{9}, 1446--1454 (2021).

\bibitem{Tein2021}
U.~Te{\u{g}}in, M.~Y{\i}ld{\i}r{\i}m, {\.{I}}.~O{\u{g}}uz, C.~Moser, and
  D.~Psaltis, \enquote{Scalable optical learning operator,}
  {\protect\JournalTitle{Nature Computational Science}} \textbf{1}, 542--549
  (2021).

\bibitem{ortin2015unified}
S.~Ort{\'\i}n, M.~C. Soriano, L.~Pesquera, D.~Brunner, D.~San-Mart{\'\i}n,
  I.~Fischer, C.~Mirasso, and J.~Guti{\'e}rrez, \enquote{A unified framework
  for reservoir computing and extreme learning machines based on a single
  time-delayed neuron,} {\protect\JournalTitle{Sci. Rep.}} \textbf{5}, 14945
  (2015).

\bibitem{xu2020}
X.~Xu, M.~Tan, B.~Corcoran, J.~Wu, T.~G. Nguyen, A.~Boes, S.~T. Chu, B.~E.
  Little, R.~Morandotti, A.~Mitchell, D.~G. Hicks, and D.~J. Moss,
  \enquote{Photonic perceptron based on a kerr microcomb for high-speed,
  scalable, optical neural networks,} {\protect\JournalTitle{Laser Photonics
  Rev.}} \textbf{14}, 2000070 (2020).

\bibitem{xu2021}
X.~Xu, M.~Tan, B.~Corcoran, J.~Wu, A.~Boes, T.~G. Nguyen, S.~T. Chu, B.~E.
  Little, D.~G. Hicks, R.~Morandotti, A.~Mitchell, and D.~J. Moss, \enquote{11
  tops photonic convolutional accelerator for optical neural networks,}
  {\protect\JournalTitle{Nature}} \textbf{589}, 44--51 (2021).

\bibitem{feldmann2021}
J.~Feldmann, N.~Youngblood, M.~Karpov, H.~Gehring, X.~Li, M.~Stappers, M.~L.
  Gallo, X.~Fu, A.~Lukashchuk, A.~S. Raja, J.~Liu, C.~D. Wright, A.~Sebastian,
  T.~J. Kippenberg, W.~H.~P. Pernice, and H.~Bhaskaran, \enquote{Parallel
  convolutional processing using an integrated photonic tensor core,}
  {\protect\JournalTitle{Nature}} \textbf{589}, 52--58 (2021).

\bibitem{Lupo2021}
A.~Lupo, L.~Butschek, and S.~Massar, \enquote{Photonic extreme learning machine
  based on frequency multiplexing,} {\protect\JournalTitle{Optics Express}}
  \textbf{29}, 28257 (2021).

\bibitem{Lupo2022}
A.~Lupo and S.~Massar, \enquote{Parallel extreme learning machines based on
  frequency multiplexing,} {\protect\JournalTitle{Applied Sciences}}
  \textbf{12} (2022).

\bibitem{yildirim2022}
M.~Yildirim, I.~Oguz, F.~Kaufmann, M.~R. Escal{\'{e}}, R.~Grange, D.~Psaltis,
  and C.~Moser, \enquote{Nonlinear optical feature generator for machine
  learning,} {\protect\JournalTitle{{APL} Photonics}} \textbf{8} (2023).

\bibitem{Butschek2022}
L.~Butschek, A.~Akrout, E.~Dimitriadou, A.~Lupo, M.~Haelterman, and S.~Massar,
  \enquote{Photonic reservoir computer based on frequency multiplexing,}
  {\protect\JournalTitle{Optics Letters}} \textbf{47}, 782 (2022).

\bibitem{Zhou2022}
T.~Zhou, F.~Scalzo, and B.~Jalali, \enquote{Nonlinear schrödinger kernel for
  hardware acceleration of machine learning,} {\protect\JournalTitle{Journal of
  Lightwave Technology}} \textbf{40}, 1308--1319 (2022).

\bibitem{Sorokina2020}
M.~Sorokina, \enquote{Multi-channel optical neuromorphic processor for
  frequency-multiplexed signals,} {\protect\JournalTitle{Journal of Physics:
  Photonics}} \textbf{3}, 014002 (2020).

\bibitem{Fischer2023}
B.~Fischer, M.~Chemnitz, Y.~Zhu, N.~Perron, P.~Roztocki, B.~MacLellan, L.~D.
  Lauro, A.~Aadhi, C.~Rimoldi, T.~H. Falk, and R.~Morandotti,
  \enquote{Neuromorphic computing via fission-based broadband frequency
  generation,} {\protect\JournalTitle{Advanced Science}}  (2023).

\bibitem{Pauwels2019}
J.~Pauwels, G.~Verschaffelt, S.~Massar, and G.~Van~der Sande,
  \enquote{Distributed kerr non-linearity in a coherent all-optical fiber-ring
  reservoir computer,} {\protect\JournalTitle{Frontiers in Physics}} \textbf{7}
  (2019).

\bibitem{Kryzhanovsky2003}
B.~Kryzhanovsky, L.~Litinskii, and A.~Fonarev, \enquote{Parametrical neural
  network based on the four-wave mixing process,}
  {\protect\JournalTitle{Nuclear Instruments and Methods in Physics Research
  Section A: Accelerators, Spectrometers, Detectors and Associated Equipment}}
  \textbf{502}, 517--519 (2003).

\bibitem{kryzhanovsky2001parametric}
B.~Kryzhanovsky, V.~Kryzhanovsky, A.~Mikaelian, and A.~Fonarev,
  \enquote{Parametric dynamic neural network recognition power,}
  {\protect\JournalTitle{Optical Memory and Neural Networks}} \textbf{10},
  211--218 (2001).

\bibitem{Agrawal}
G.~Agrawal, \emph{Nonlinear Fiber Optics, Fifth ed.} (Elsevier, 2013).

\bibitem{Jonuzi2023A}
T.~Jonuzi, A.~Lupo, M.~C. Soriano, S.~Massar, and J.~D. Domen{\'{e}}ch,
  \enquote{Integrated programmable spectral filter for frequency-multiplexed
  neuromorphic computers,} {\protect\JournalTitle{Optics Express}} \textbf{31},
  19255 (2023).

\bibitem{Jonuzi2023}
T.~Jonuzi, A.~Lupo, M.~C. Soriano, J.~D.~D. Gomez, and S.~Massar,
  \enquote{Integrated optical output layer for a reservoir computer based on
  frequency multiplexing,} in \emph{{AI} and Optical Data Sciences {IV},}
  K.~ichi Kitayama and B.~Jalali, eds. ({SPIE}, 2023).

\bibitem{heart_dataset}
A.~Janosi, W.~Steinbrunn, M.~Pfisterer, and R.~Detrano, \enquote{Heart diseases
  data set,} \url{https://archive.ics.uci.edu/ml/datasets/heart+disease}.

\bibitem{wine_dataset}
M.~Forina, \enquote{Wine data set,}
  \url{https://archive.ics.uci.edu/ml/datasets/wine}.

\bibitem{iris_dataset}
R.~Fisher, \enquote{Iris data set,}
  \url{http://archive.ics.uci.edu/ml/datasets/iris}.

\bibitem{Yeh2018}
I.-C. Yeh and T.-K. Hsu, \enquote{Building real estate valuation models with
  comparative approach through case-based reasoning,}
  {\protect\JournalTitle{Applied Soft Computing}} \textbf{65}, 260--271 (2018).

\bibitem{Balac16}
S.~Balac, A.~Fernandez, F.~Mah{\'{e}}, F.~M{\'{e}}hats, and R.~Texier-Picard,
  \enquote{The interaction picture method for solving the generalized nonlinear
  schr\"{o}dinger equation in optics,} {\protect\JournalTitle{{ESAIM}:
  Mathematical Modelling and Numerical Analysis}} \textbf{50}, 945--964 (2016).

\bibitem{zajnulina_2015OFC}
M.~Zajnulina, J.~M. Chavez~Boggio, M.~Böhm, A.~A. Rieznik, T.~Fremberg,
  R.~Haynes, and M.~M. Roth, \enquote{Generation of optical frequency combs via
  four-wave mixing processes for low- and medium-resolution astronomy,}
  {\protect\JournalTitle{Applied Physics B}} \textbf{120}, 171–184 (2015).

\bibitem{Liu2022}
Y.~Liu, Z.~Qiu, X.~Ji, A.~Lukashchuk, J.~He, J.~Riemensberger, M.~Hafermann,
  R.~N. Wang, J.~Liu, C.~Ronning, and T.~J. Kippenberg, \enquote{A photonic
  integrated circuit{\textendash}based erbium-doped amplifier,}
  {\protect\JournalTitle{Science}} \textbf{376}, 1309--1313 (2022).

\bibitem{zajnulina_soa2017}
M.~Zajnulina, B.~Lingnau, and K.~Lüdge, \enquote{Four-wave mixing in
  quantum-dot semiconductor optical amplifiers: A detailed analysis of the
  nonlinear effects,} {\protect\JournalTitle{IEEE Journal of Selected Topics in
  Quantum Electronics}} \textbf{23}, 1--12 (2017).

\bibitem{Gaeta2019}
A.~L. Gaeta, M.~Lipson, and T.~J. Kippenberg, \enquote{Photonic-chip-based
  frequency combs,} {\protect\JournalTitle{Nature Photonics}} \textbf{13},
  158--169 (2019).

\end{thebibliography}
\end{document}